\documentclass[sigconf]{acmart}

\usepackage{graphicx}
\usepackage{amsmath,amsfonts,amssymb}
\usepackage{dsfont}
\usepackage{enumitem}
\usepackage{epstopdf}
\usepackage{algorithm}
\usepackage{algpseudocode}
\usepackage{url}
\usepackage{color}
\usepackage{changepage}
\usepackage[font=small]{subcaption}

\newcommand{\ignore}[1]{}
\begin{document}

\title{Talking to Your TV: Context-Aware Voice Search\\ with Hierarchical Recurrent Neural Networks}

\author{Jinfeng Rao$^{1,2}$, Ferhan Ture$^{1}$, Hua He$^{2}$, Oliver Jojic$^{1}$, and Jimmy Lin$^{3}$}
\affiliation{\vspace{0.1cm}
\institution{$^{1}$ Comcast Labs}
\institution{$^{2}$ Department of Computer Science, University of Maryland}
\institution{$^{3}$ David R. Cheriton School of Computer Science, University of Waterloo}
}
\email{jinfeng@cs.umd.edu, ferhan_ture@comcast.com}

\begin{abstract}
We tackle the novel problem of navigational voice queries posed
against an entertainment system, where viewers interact with a
voice-enabled remote controller to specify the program to watch. This
is a difficult problem for several reasons:\ such queries are short,
even shorter than comparable voice queries in other domains, which
offers fewer opportunities for deciphering user intent.  Furthermore,
ambiguity is exacerbated by underlying speech recognition errors. We
address these challenges by integrating word- and character-level
representations of the queries and by modeling voice search sessions
to capture the contextual dependencies in query sequences.  Both are
accomplished with a probabilistic framework in which recurrent and
feedforward neural network modules are organized in a hierarchical
manner. From a raw dataset of 32M voice queries from 2.5M viewers on
the Comcast Xfinity X1 entertainment system, we extracted
data to train and test our models. We demonstrate the benefits of our
hybrid representation and context-aware model, which significantly
outperforms models without context as well as the current deployed
product.

\end{abstract}

\maketitle

\section{Introduction}

\noindent Voice-based interactions with computing devices are becoming
increasingly prevalent, driven by several convergent trends. The
ubiquity of smartphones and other mobile devices with restrictive
input methods makes voice an attractive modality for
interaction:\ Apple's Siri, Microsoft's Cortana, and the Google
Assistant are prominent examples. Google observed that there are more
searches taking place from mobile devices than from traditional
desktops,\footnote{\small \url{http://selnd.com/1c1tKXg}} and that
20\% of mobile searches are voice queries.\footnote{\small Stated by
  Google CEO Sundar Pichai during Google I/O 2016.} The success of
these products has been enabled by advances in automatic speech
recognition (ASR), thanks mostly to deep learning.

Increasing comfort with voice-based interactions, especially with
AI-agents, feeds into the emerging market on ``smart homes''. Products
such as Amazon Echo and Google Home allow users to control a variety
of devices via voice (e.g., ``turn on the TV'', ``play music by
Adele''), and to issue voice queries (e.g., ``what's the weather
tomorrow?''). The market success of these products demonstrates that
people do indeed want to control smart devices in their environment
via voice.

In this paper, we tackle the problem of navigational voice queries
posed against an entertainment system, where viewers interact with a
voice-enabled remote controller to specify the program (TV shows,
movies, sports games)\ they wish to watch. If a viewer wishes to watch
the popular series ``Game of Thrones'', saying the name of the program
should switch the television to the proper channel. This is simpler
and more intuitive than scrolling through channel guides or awkwardly
trying to type in the name of the show on the remote controller. Even
if the viewer knows that Game of Thrones is on HBO, finding the right
channel may still be challenging, since entertainment packages may
have hundreds of channels.

Our problem is challenging for a few reasons. Viewers have access to
potentially tens of thousands of programs, especially if we include
on-demand titles. Program names can be highly ambiguous. For instance,
the query ``Chicago Fire'' could refer to either the television series
or a soccer team. Even with recent advances, ASR errors can exacerbate
the ambiguity by transcribing queries like ``Caillou'' (a Canadian
children's education television series) as ``you''. Based on our
analysis of 32M voice queries in this domain, we find that they are
shorter (average of 2.04 words) than published statistics about voice
queries on smartphones and computers~\cite{guy2016searching,
  schalkwyk2010your}. Short queries make the prediction problem more
difficult because there is less signal to extract.

\smallskip \noindent {\bf Contributions.} We tackle the above
challenges using two key ideas to infer user intent:\ hybrid query
representations and modeling search sessions. Specifically, our
contributions are as follows:

\begin{itemize}[leftmargin=*]

\item To our knowledge, we are the first to systematically study voice
  queries in the entertainment context. We propose a technique to
  automatically collect ground truth labels for voice query sessions
  from real-world usage data by examining viewing behaviors following
  the sessions.

\item Our probabilistic model has two key features:\ First, we
  integrate word- and character-level representations of the
  queries. Second, we model voice search sessions to understand the
  contextual dependencies in query sequences. Both are accomplished
  with a probabilistic framework in which recurrent and feedforward
  neural network modules are organized in a hierarchical manner.

\item Evaluations on a large real-world dataset demonstrate the
  effectiveness of our hybrid query representation and context-aware
  models, significantly outperforming strong baselines as well as the
  current deployed system. Detailed analyses clarify {\it how} our
  models are better able to understand user intent.

\end{itemize} 

\section{Background and Related Work}

The context of our work is voice search on the Comcast Xfinity X1
entertainment platform, by one of the largest cable companies in the
United States with approximately 22 million subscribers in 40
states. X1 is essentially a software package distributed as part of
the X1 cable box, which has been deployed to 17 million customers
since around 2015. X1 can be controlled via the ``voice remote'',
which is a remote controller that has an integrated microphone to
receive voice queries from viewers. The current deployed system is
based on a combination of hand-crafted rules and machine-learned
models to arrive at a final response. The system has a diverse set of
capabilities, which increases query ambiguity and magnifies the
overall challenge of understanding user intent. These capabilities
range from channel change to entity search (e.g., sports team, person,
movie, etc.). In addition, voice queries may involve general questions,
from home security control to troubleshooting the wifi network, or may
be ultimately directed to external apps such as Pandora. In this
paper, we focus on navigational voice queries where viewers specify
the program they wish to watch.

In our particular application, we receive as input the one-best result
of the ASR system, which is a text string. We do not have access to
the acoustic signal, as the ASR system is a black box. Although it
would be ideal if we could build joint models over both the acoustic
signals, transcription lattice, and user intent, in many operational
settings this is not practical or even possible. In the case of X1,
for a variety of reasons, the ASR is outsourced to a third party---a
scenario not uncommon in many organizations who do not wish to invest
in ASR from the ground up. Thus, as we described in the introduction,
transcription error compounds the ambiguity in the queries and
introduces additional complexity that our models need to handle.

We are, of course, not the first to tackle voice
search~\cite{wang2008introduction,
  acero2008live,feng2009effects,chelba2013empirical,shan2010search},
although to our knowledge we are the first to focus on voice queries
directed at an entertainment system. How is this particular domain
different?  The setting is obviously different---in our case, viewers
are clearly sitting in front a television with an entertainment
intent. To compare and contrast viewers' actual utterances, we can
turn to previously-published work that studied the characteristics of
voice search logs, especially in comparison to text search
data~\cite{crestani2006written,yi2011mobile,guy2016searching,
  schalkwyk2010your}. Schalkwyk et al.~\cite{schalkwyk2010your}
reported statistics of queries collected from Google Voice search
logs, which found that short queries, in particular 1-word and 2-word
queries, were more common in the voice search setting, while long queries
were much rarer. In contrast, in a more recent study,
Guy~\cite{guy2016searching} reported that voice queries tend to be
longer than text queries, based on a half-million query dataset from
the Yahoo mobile search application. In addition, Guy studied the
characteristics of voice queries in a more comprehensive way,
including query term frequencies, popularity, syntax, post-click
behaviors, etc. The average length across 32M voice queries is 2.04
in our dataset, much shorter than the reported average of 4.2 for
Yahoo voice search\footnote{Similar conclusions follow for other
  length-based statistics:\ median was 2 (vs.\ 4), maximum was 69
  (vs.\ 109), and standard deviation was 1.23
  (vs.\ 2.96).}~\cite{guy2016searching}.

We note another important difference between our entertainment context
and voice search applications on smartphones:\ on a mobile device, it
is common to back off to a web search if the query intent is not
identified with high confidence. For Yahoo, Guy reported that less
than half of voice queries (43.3\%) are handled by a pre-defined
card. While we are not aware of any scientific study about web
browsing behavior on a TV, our intuition is that a list of search
results is less useful to TV viewers than it might be for smartphone
users, since subsequent interactions are much more awkward:\ it is
difficult for users to scroll and they have limited input methods for
follow-up interactions. Furthermore, televisions are not optimized for
browsing webpages at a distance.

Personalization can help disambiguate
queries~\cite{zweig2011personalizing,bennett2012modeling}, since
user preference is an important signal in deciphering user
intent. However, since the TV is usually shared amongst the household,
the feasibility of reliable personalization is not as clear as on a
smartphone or computer (i.e., not obvious low-hanging fruit).  This
makes it even more important to exploit other signals.

There is also research on voice query reformulations that is relevant
to our work on modeling sessions~\cite{jiang2013users,
  hassan2015characterizing, shokouhi2014mobile, shokouhi2016did}. For
example, Jiang et al.~\cite{jiang2013users} analyzed different types
of voice recognition errors and users' corresponding reformulation
strategies. Hassan et al.~\cite{hassan2015characterizing} built
classifiers to differentiate between reformulated and non-reformulated
query pairs. The study by Shokouhi et al.~\cite{shokouhi2014mobile}
suggested that users don't prefer to switch between voice and text
when reformulating queries. A more recent
paper~\cite{shokouhi2016did} proposed an automatic way to label voice
queries by examining post-click and reformulation behaviors, which
produced a large amount of ``free'' training data to reduce ASR
errors. These papers provide a source of inspiration for our models.

Our approach to tackling the challenges associated with ambiguous
voice queries is to take advantage of context. Our fundamental
assumption is that when the viewer is not satisfied with the results
of a query, she will issue another query in rapid succession and
continue until the desired program is found or until she gives
up. Note that these sequences often represent refinement of user
intent:\ part of the process is the viewer deciding what to watch.  By
modeling voice search sessions (i.e., sequences of successive voice
queries), we can better understand the viewer's underlying intent. For
example, compare two sessions:\ [``tv shows'', ``ncis'', ``cargo
  fire'', ``chicago fire''] and [``espn'', ``chicago sports'',
  ``chicago fire''].  Although both end in the same query, it is
fairly clear that in the first case, the viewer is interested in the
TV drama series ``Chicago Fire'' (since previous queries all mention
other drama series), whereas in the second session, it is clear that
the viewer is interested in the sports team with the same name. This
idea, of course, is not novel, and there is a large body of literature
focused on exploiting web search
sessions~(e.g.,~\cite{jones2008beyond,cao2009context,bennett2012modeling,guan2013utilizing,liu2010personalizing,
  luo2015session, zhang2015information}, just to mention a few). A
comprehensive survey is beyond the scope of this paper, but previous
work is concerned with text-based web search, which differs both in
modality and in domain.

\section{Model Architecture}

\subsection{Problem Formulation}

Given a voice query session $[q_1, \ldots, q_n]$, our task is to
predict the program $p$ that the user intends to watch.  We perform
this prediction cumulatively at each time step $t \in [1, n]$ on each
successive new voice query $q_t$, exploiting all previous queries in
the session, $[q_1, \ldots , q_{t-1})$. For example, consider a
three-query session $s_i=[q_{i_1}, q_{i_2}, q_{i_3}]$, there will be
three separate predictions:\ first with $[q_{i_1}]$, second with
$[q_{i_1}, q_{i_2}]$, and third with $[q_{i_1}, q_{i_2}, q_{i_3}]$.
We sessionize the voice query logs heuristically based on a time gap
(in this case, 45 seconds---more details later), similar to how web
query logs are sessionized based on inactivity.  As described above,
each query is a text string, the output of a third-party ``black box''
ASR system that we do not have internal access to.

We aim to learn a mapping function $\Theta$ from a query sequence to a
program prediction, modeled using a probabilistic framework:

\begin{align*}
\textrm{Data: } D &= \{(s_i, p_i)\ |\ s_i = [q_{i_1}, ..., q_{i_{|s_i|}}],\ p_i \in \Phi\}_1^{|D|}\\
\textrm{Model: } \hat{\theta} &= \arg \max_\theta \prod_{i=1}^{|D|} \prod_{t=1}^{|s_i|} P(p_i | q_{i_1}, ... , q_{i_t}; \theta)
\label{eq:problem}
\end{align*}

\noindent where $D$ denotes a set of labeled sessions ($s_i$ denotes
the $i$-th session with $|s_i|$ queries), $p_i$ is the intended
program for session $i$, $\Phi$ is the global set of programs, and
$\theta$ is the set of parameters in the mapping function
$\Theta$. Our goal is to maximize the product of prediction
probabilities.

We decompose the program prediction task into learning three mapping
functions: a query embedding function
$\mathds{F}(x;\theta_{\mathds{F}})$, a contextual function
$\mathds{G}(x; \theta_{\mathds{G}})$, and a classification function
$\mathds{H}(x; \theta_{\mathds{H}})$. The query embedding function
$\mathds{F}(\cdot)$ takes the text of the query as input and produces
a semantic representation of the query; the contextual function
$\mathds{G}(\cdot)$ considers representations of all the preceding
queries as context and maps them to a high-dimensional embedding
vector to capture both semantic and contextual features; finally, the
classification function $\mathds{H}(\cdot)$ predicts possible programs
from the learned contextual vector. We adopt the following
decomposition:

\begin{equation}
\begin{split}
P(p_i | q_{i_1}, ... , q_{i_t}) \sim\ &P(p_i | c_{i_t}) \cdot P(c_{i_t} | v_{i_1}, ... , v_{i_t}) \\
&\cdot P(v_{i_1}, ... , v_{i_t} | q_{i_1}, ... , q_{i_t})
\end{split}
\label{eq:decompose}
\end{equation}

\noindent where $c_{i_t}$ denotes the \emph{contextual} embedding of
the first $t$ queries in the $i\textrm{-th}$ session and $v_{i_t}$
denotes the embedding of the $t\textrm{-th}$ query of the
$i\textrm{-th}$ session. The relationship between these embeddings can
be formulated using the three mapping functions above:\ $\mathds{F}$
maps a query $q_{i_j}$ to its embedding $v_{i_j}$ in vector space;
$\mathds{G}$ maps a sequence of query embeddings $[v_{i_1}, ... ,
  v_{i_t}]$ to a contextual embedding $c_{i_t}$; and $\mathds{H}$ maps
the contextual embedding to a program $p_i$:

\begin{equation*}
\begin{split}
v_{i_t} & \sim \mathds{F}(q_{i_t}; \theta_{\mathds{F}}) \\
c_{i_t} & \sim \mathds{G}(v_{i_1}, ... , v_{i_t}; \theta_{\mathds{G}}) \\
p_i & \sim \mathds{H}(c_{i_t}; \theta_{\mathds{H}}) \\
& 1 \le t \le |s_i|
\end{split}
\end{equation*}

\noindent By assuming that each query is embedded independently, we
can reduce the last term in Equation~(\ref{eq:decompose}) as follows:

\begin{equation*}
\begin{split}
P(p_i | q_{i_1}, ... , q_{i_t})= P(p_i | c_{i_t}) \cdot P(c_{i_t} | v_{i_1}, ... , v_{i_t})
\cdot \prod_{j=1}^t P(v_{i_j} | q_{i_j})
\end{split}
\end{equation*}

\noindent We model the query embedding function $\mathds{F}(\cdot)$
and the contextual function $\mathds{G}(\cdot)$ by organizing two Long
Short-Term Memory (LSTM)~\cite{hochreiter1997long} models in a
hierarchical manner. The decision function $\mathds{H}(\cdot)$ is
represented as a feedforward neural network layer. Before we introduce
the details of our model architecture, we provide an overview of the
LSTM model.

\subsection{Long Short-Term Memory Networks}

Long Short-Term Memory (LSTM)~\cite{hochreiter1997long} networks are
well-known for being able to capture long-range contextual
dependencies over input sequences. This is accomplished by using a
sequence of memory cells to store and memorize historical information,
where each memory cell contains three gates (input gate, forget gate,
and output gate) to control the information flow. The gating mechanism
enables the LSTM to handle the gradient vanishing/explosion problem
for long sequences of inputs.

Given an input sequence $\textbf{x} = (x_1, ... , x_T)$, an LSTM model
outputs a sequence of hidden vectors $\textbf{h} = (h_1, ... ,
h_T)$. A memory cell at position $t$ digests the input element $x_t$
and previous state information $h_{t-1}$ to produce updated state
$h_t$ as follows:

\begin{equation*}
\begin{split}
i_t & = \sigma(W_{xi} x_t + W_{hi} h_{t-1} + b_i) \\
f_t & = \sigma(W_{xf} x_t + W_{hf} h_{t-1} + b_f) \\
o_t & = \sigma(W_{xo} x_t + W_{ho} h_{t-1} + b_o) \\
c_t & = f_t \cdot c_{t-1} + i_t \cdot \sigma(W_{xc} x_t + W_{hc} h_{t-1} + b_c) \\
h_t & = o_t \cdot \tanh(c_t)
\end{split}
\end{equation*}

\noindent where the $W$ terms are weight matrices, the $b$ terms
represent bias vectors, $\sigma$ is the sigmoid activation function,
and $i$, $f$, $o$, and $c$ are respectively the input gate, forget
gate, output gate, and cell vectors, with each having the same size as
the output vector $h$. In this paper, we refer to the size of the
output vector $h$ as the LSTM size.

In many application scenarios, the input sequence $\textbf{x}$ can
vary in length for different instances (i.e., queries can have
different numbers of words and characters in our task). There are two
standard ways to handle this variable length issue. One way is to perform
an initial scan over a single batch or the entire dataset to obtain the
maximum sequence length, then create an array of memory cells with the
maximum length. Whenever a sequence element $x_t$ arrives, the memory
cell at index $t$ will digest the input element and produce the hidden
state $h_t$. The other way is to dynamically allocate space for
storing new memory cells only when the arriving instance $\textbf{x}$
has a greater length than all previous instances. The created LSTM
memory cells all share the same parameters. We use the second strategy
(what we call \emph{dynamic allocation policy}) in our implementations
to avoid needing an initial scan.

\subsection{Query Representation}\label{sec:query}

Since query strings serve as the sole input in our model, an
expressive query representation is essential to accurate predictions.
We represent each query as a sequence of elements (words or
characters); each element is passed through a lookup layer and
projected into a $d$-dimensional vector, thereby representing the
query as an $m\times d$ matrix ($m$ is the number of elements in the
query). We consider three variations of this representation:

\begin{enumerate}[leftmargin=*]

\item \textbf{Character-level} representation, which encodes a query
  as a sequence of characters and the lookup layer converts each
  character to a one-hot vector. In this case, $m$ would be the number
  of characters in the query and $d$ would be the size of the character
  dictionary of the entire dataset.

\item \textbf{Word-level} representation, which encodes the query as a
  sequence of words, and the word vectors are read from a pre-trained
  word embedding, e.g., word2vec~\cite{Mikolov:2013aa}. In this case,
  $d$ would be the dimensionality of the word embedding.

\item \textbf{Combined} representation, which combines both the
  character-level and word-level representations by feeding the
  representations to two separate query embedding functions
  $\mathds{F}_c$ and $\mathds{F}_w$, respectively, then concatenating
  the two learned vectors $v_c$ and $v_w$ as the combined query
  embedding vector.

\end{enumerate}

\noindent Our intuition for these different representations is as
follows:\ Based on our analysis of voice query logs, we observe
many unsatisfactory responses due to speech recognition errors. For
example, voice queries intended for the program ``Caillou'' (a
Canadian children's education television series) are often recognized
as ``Cacio'' or ``you''. Capturing such variations with a word-level
representation would likely suffer from data sparsity issues. On the
other hand, initializing a query through word embedding vectors would
encode words in a semantic vector space, which would help in matching
queries to programs based on semantic relatedness (e.g., the query
``Portland Trail Blazers'' is semantically similar to the intended
program ``NBA basketball'' without any words in common). Word embeddings
are also useful for recognizing semantically-similar contextual clues
such as ``Search'', ``Find'' or ``Watch''. With a character-level
representation, such similarities would need to be learned from
scratch. Whether the benefits of either representation balance the
drawbacks is an empirical question we study through experiments, but
we hypothesize that a combined representation would capture the best
of both worlds.

\subsection{Basic Model}\label{sec:basic}

\begin{figure}[t]
\centering
\includegraphics[width=0.5\linewidth]{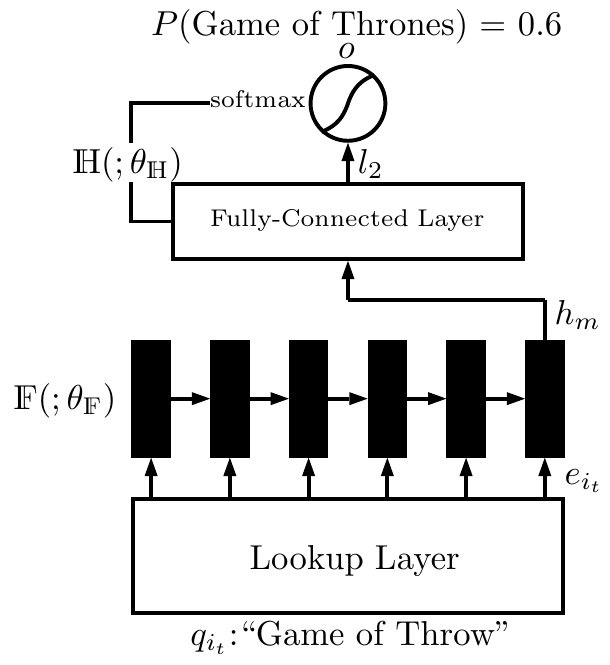}
\vspace{-0.3cm}
\caption{Architecture of the Basic Model} \label{independent}
\end{figure}

In the basic context-independent model, queries in a session are
assumed to be independent and thus we do not attempt to model context.
That is, each query is treated as a complete sample for model
inference and prediction. The mapping function $\Theta$ from query to
program can be simplified as follows:

\begin{equation}
\begin{split}
\Theta & \sim \arg \max_\theta \prod_{i=1}^{|D|} \prod_{t=1}^{|s_i|} P(p_i | q_{i_t}) \\
& = \arg \max_\theta \prod_{i=1}^{|D|} \prod_{t=1}^{|s_i|} P(p_i | v_{i_t}) P(v_{i_t} | q_{i_t})
\end{split}
\label{equation:independent}
\end{equation}
\begin{equation*}
v_{i_t} \sim \mathds{F}(q_{i_t}; \theta_{\mathds{F}}), \quad p_i \sim \mathds{H}(v_{i_t}; \theta_{\mathds{H}}), \quad 1 \le t \le |s_i|
\end{equation*}

\noindent Here, the program $p_i$ is only dependent on the current
query $q_{i_t}$. The contextual function $\mathds{G}(\cdot)$ is
modeled as an identity function since there is no context from our
assumption.

The architecture of the basic model is shown in
Figure~\ref{independent}. In the bottom, we use an LSTM as our query
embedding function $\mathds{F}(\cdot)$. The text query is projected
into an $m\times d$ dimensional matrix through the lookup layer, then
fed to the LSTM, which has $m$ memory cells and each cell processes an
element vector. The hidden state at the last time step $h_m$ is used
as the query embedding vector $v$.  At the top, there is a
fully-connected layer followed by a soft-max layer for learning the
classification function $\mathds{H}(\cdot)$. The fully-connected layer
consists of two linear layers with one element-wise activation layer
in between. Given the query embedding vector $v$ as input, the
fully-connected layer computes the following:

\begin{equation*}
\begin{split}
&l = \sigma(W_{h_1} \cdot v + b_{h_1}) \\
&l_2 = W_{h_2} \cdot l + b_{h_2}
\end{split}
\end{equation*}

\noindent where the $W$ terms are the weight matrices and the $b$
terms are bias vectors. We use the $\textrm{tanh}$ function as the
non-linear activation function $\sigma$, which is commonly adopted in
many neural network architectures. The soft-max layer normalizes the
vector $l_2$ to a $L1$ norm vector $o$, with each output score
$o[p_j]$ denoting the probability of producing program $p_j$ as
output:

\begin{equation*}
o[p_j] = \frac{\exp(l_2[p_j] - \textrm{shift})}{\sum_{p_k=1}^{|\Phi|} \exp(l_2[p_k] - \textrm{shift})}
\end{equation*}
\noindent where $\textrm{shift} = \max_{p_k=1}^{|\Phi|} l_2[p_k]$, $|\Phi|$ is the total number of programs in the dataset. 

\begin{algorithm}[t]
\small
\begin{algorithmic}[1]
      \For{each session $s_i$ in the dataset $i= 1 ... |D|$}
        \For{each query $q_{i_t}$ in session $s_i$ with $t= 1 ... |s_i|$}
        \State \Comment{Forward Prediction Start}
        	\State $e_{i_t} = \textrm{encode}(q_{i_t})$ \label{line:encode}
         \State $h_{1,...,m} = \textrm{LSTM:forward}(e_{i_t})$ \label{line:lstm}
         \State $l_2 = \textrm{FC:forward}(h_m)$ \label{line:fc}
         \State $o = \textrm{softmax:forward}(l_2)$
          \State $loss = \textrm{criterion:forward}(o, p_i)$  
          \State  \Comment{Backward Propagation Start}
          \State $\textrm{grad$\_$criterion} = \textrm{criterion:backward}(o, p_i)$
          \State $\textrm{grad$\_$soft} = \textrm{softmax:backward}(l_2, \textrm{grad$\_$criterion})$
          \State $\textrm{grad$\_$linear} = \textrm{FC:backward}(h_m, \textrm{grad$\_$soft})$
          \State $\textrm{grad$\_$lstm} = \textrm{zeros}(m, \textrm{lstm$\_$size})$ \label{line:gradlstm1}
          \State $\textrm{grad$\_$lstm}[m] = \textrm{grad$\_$linear}$
          \State $\textrm{LSTM:backward}(e_{it}, \textrm{grad$\_$lstm})$ \label{line:gradlstm2}
          \State $\textrm{update$\_$parameters}()$ \label{line:updatesimple}
        \EndFor
      \EndFor   
\end{algorithmic}
\caption{Training the Basic Model}\label{alg:basic}
\end{algorithm}

We adopt the negative log likelihood loss function to train the model,
which is derived from Equation~(\ref{equation:independent}):

\begin{equation*}
\begin{split}
L & = -\sum_{i=1}^{|D|} \sum_{t=1}^{|s_i|} \log P(p_i | q_{i_t}) + \lambda \cdot \|<\theta_\mathds{F}, \theta_\mathds{G}, \theta_\mathds{H}>\|^2 \\
&= -\sum_{i=1}^{|D|} \sum_{t=1}^{|s_i|} \log o_{i_t}[p_i] + \lambda \cdot \|<\theta_\mathds{F}, \theta_\mathds{G}, \theta_\mathds{H}>\|^2
\end{split}
\end{equation*}

\noindent where $o_{i_t}$ is the score vector computed from query
$q_{i_t}$ and $p_i$ is the true program for session $i$; $\lambda$ is
the regularization weight and $<\theta_\mathds{F}, \theta_\mathds{G},
\theta_\mathds{H}>$ is the set of model parameters. The optimization
goal is to minimize the loss criterion $L$.

The training process is shown in Algorithm~\ref{alg:basic}. The
overall structure is to iterate over each query in all sessions to
perform the forward prediction and backward propagation
operations. The \emph{forward} phase follows our model architecture in
Figure~\ref{independent}. A query is first encoded as a matrix in
Line~\ref{line:encode} by specifying the encoding method (i.e.,
character, word, or combined). Line \ref{line:lstm} computes the output
states $h$ by feeding an input matrix to the LSTM model. FC in
Line~\ref{line:fc} denotes the fully-connected layer. In the
\emph{backward} phase, each module requires the original inputs and
the gradients propagated from its upper layer to compute the gradients
with respect to the inputs and its own parameters.  It is worth noting that in
Lines~\ref{line:gradlstm1}-\ref{line:gradlstm2} the gradients
$\textrm{grad$\_$lstm}$ are initialized to zero for the first $m-1$
cells.  This is because in the forward phase, we only use the last
LSTM state $h_m$ as the query embedding vector for the upper layers
and throw away the other states $h_{1, ...,
  m-1}$. Line~\ref{line:updatesimple} performs gradient descent to
update model parameters.  All the forward and backward functions used
here are written as black box operations, and we refer interested
readers elsewhere~\cite{Goodfellow-et-al-2016} for more details.

\subsection{Full Context Model}\label{sec:context}

\begin{figure}[t]
\centering
\includegraphics[width=0.9\linewidth]{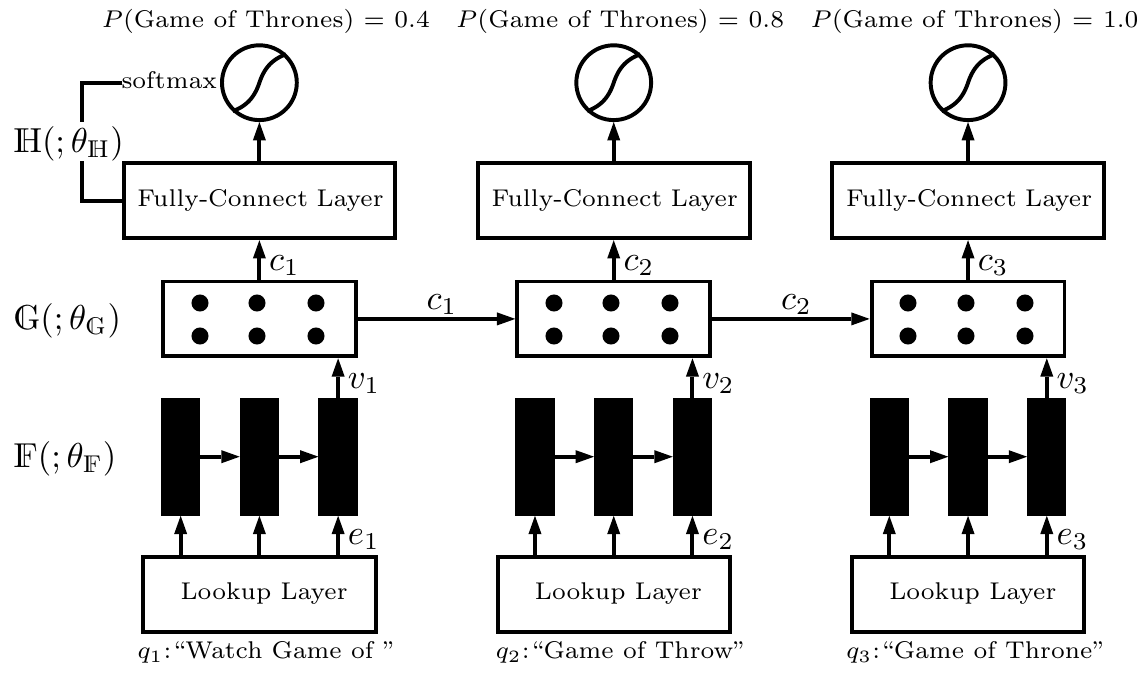}
\caption{Architecture of the Full Context Model}
\label{fig:context}
\end{figure}

We propose two approaches to modeling context:\ the full context model
(presented here) and the constrained context model (presented next).
The architecture of the full context model is shown in
Figure~\ref{fig:context}, which uses the basic model as a building
block.  We use another LSTM (the dotted rectangle in the middle of
Figure~\ref{fig:context}) to learn the contextual function
$\mathds{G}(v_1, ..., v_t; \theta_{\mathds{G}})$. Previous query
embedding vectors $[v_1, ..., v_{t-1}]$ are encoded as a context
vector $c_{t-1}$, which is combined with the current query embedding
vector $v_t$ and fed to the LSTM memory cell at time $t$.  This allows
the LSTM to find an optimal combination of signals from the previous
context and the current query. For sessions with a single underlying
intent (i.e., the user is consistently looking for a specific
program), the model can learn the intrinsic relatedness between
successive queries and continuously reinforce confidence in the true
intent. In reality, context can sometimes be irrelevant (e.g., user
zapping through channels), which might introduce noise. When the
context diverges too much from the current query embeddings, the model
should be able to ignore the noisy signals to reduce their negative
impact.

We adopt a many-to-many hierarchical architecture. A query embedding
$\mathds{F}(\cdot)$ and classification layer $\mathds{H}(\cdot)$ is
applied over each query for program prediction at each time step
$t$. We hope to find the true user intent as early as possible to
reduce the interactions between the user and our voice product. The
parameters of the query embedding layer $\mathds{F}(\cdot)$, as well
as the classification layer $\mathds{H}(\cdot)$, are shared by all
queries regardless of their position in the session. For instance, two
identical queries with different positions in a session will have the
same query embedding vector. Except for the contextual layer
$\mathds{G}(\cdot)$, all other modules (e.g., query embedding,
fully-connected, soft-max layers, loss function) remain same as in the
basic (context-independent) model.

\begin{algorithm}[t]
\small
\begin{algorithmic}[1]
      \For{each session $s_i$ in the dataset $i= 1 ... |D|$}
        \State $v = \textrm{zeros}(|s_i|, \textrm{lstm$\_$size})$ \Comment{query embedding vectors} \label{line:fwd1}
        \For{each query $q_{i_t}$ with $t= 1 ... |s_i|$}
         	\State $e_{i_t} = \textrm{encode}(q_{i_t})$
         	\State $h_{1,...,m} = \textrm{LSTM[t]:forward}(e_{i_t})$ \label{line:lstmarray} 
         	\State $v_t = h_m$ \label{line:fwdlstm}
        \EndFor	\label{line:fwd2}
        \State $c_{1, ..., |s_i|} = \textrm{C$\_$LSTM:forward}(v_{1, ..., |s_i|})$  \Comment{contextual vectors} \label{line:fwdcontext}
        \State $\textrm{session$\_$loss} = 0 $	\label{line:fwdbwd1}
        \State $\textrm{grad$\_$linear} = \textrm{zeros}(|s_i|, \textrm{lstm$\_$size})$ 
        \For{each query $q_{i_t}$ with $t= 1 ... |s_i|$}
         	\State $l_2 = \textrm{FC:forward}(c_t)$
          	\State $o = \textrm{softmax:forward}(l_2)$
         	\State $\textrm{loss} = \textrm{criterion:forward}(o, p_i)$  
          	\State $\textrm{session$\_$loss} = \textrm{session$\_$loss} + loss $ \label{line:sessionloss}
          	\State $\textrm{grad$\_$criterion} = \textrm{criterion:backward}(o, p_i)$
          	\State $\textrm{grad$\_$soft} = \textrm{softmax:backward}(l_2, \textrm{grad$\_$criterion})$
          	\State $\textrm{grad$\_$linear}[t] = \textrm{FC:backward}(c_t, \textrm{grad$\_$soft})$
          \EndFor \label{line:fwdbwd2}
          \State $\textrm{grad$\_$context} = \textrm{C$\_$LSTM:backward}(v_{1, ..., |s_i|}, \textrm{grad$\_$linear})$ \label{line:bwdcontext1}
	  \For{each query $q_{i_t}$ with $t= 1 ... |s_i|$} \label{line:bwdembed}
          	\State $\textrm{grad$\_$lstm} = \textrm{zeros}(m, \textrm{lstm$\_$size})$
          	\State $\textrm{grad$\_$lstm}[m] = \textrm{grad$\_$context}[t]$
          	\State $\textrm{LSTM[t]:backward}(e_{i_t}, \textrm{grad$\_$lstm})$
        \EndFor \label{line:bwdcontext2}
        \State $\textrm{update$\_$parameters}()$ \label{line:update}
      \EndFor   
\end{algorithmic}
\caption{Training the Full Context Model}\label{alg:context}
\end{algorithm}

The training process for this model (Algorithm~\ref{alg:context})
starts with forward predictions for multiple queries in the session
(Lines~\ref{line:fwd1}-\ref{line:fwd2}). Similar to
Algorithm~\ref{alg:basic}, only the last LSTM state $h_m$ is selected
as the query embedding vector (Line~\ref{line:fwdlstm}). Since
sessions can have a variable number of queries, we use the dynamic
allocation policy to create a list of LSTMs with each LSTM ingesting a
query (i.e., LSTM[t] in
Line~\ref{line:lstmarray}). Line~\ref{line:fwdcontext} utilizes
another LSTM model to compute the context from sequential query
embeddings. Lines~\ref{line:fwdbwd1}-\ref{line:fwdbwd2} perform
forward predictions and backward propagations for multiple queries in
the classification layer. The queries are processed in a sequential
manner such that for each query all forward operations are immediately
followed by all backward operations before moving to the next
query. Lines~\ref{line:bwdcontext1}-\ref{line:bwdcontext2} propagate
the gradients through the contextual and embedding
LSTMs. Line~\ref{line:update} updates model parameters for each
session by optimizing the session loss in Line~\ref{line:sessionloss}.

It is important to note that the prediction task is applied at the
query level:\ our model tries to predict the program after {\it each}
query in the session.  The alternative is to optimize for program
prediction given {\it all} queries in the session---this is a much
easier task, since the entire session has been observed.  It also
defeats the purpose of our setup since we wish to satisfy viewer
intents as soon as possible.

\subsection{Constrained Context Model}\label{sec:c-context}

In addition to the full context model described above, we explore a
variant that we call the \emph{constrained} context model.  The model
architecture is the same as the full context model
(Figure~\ref{fig:context}).  The difference, however, lies in how we
learn the model. For the constrained context model, we adopt a
pre-training strategy as follows:\ we first train the basic model
(Algorithm~\ref{alg:basic}) and then use the learned LSTM parameters
to initialize the constrained context model's query embedding
layer. The embedding layer is then fixed and purely used for
generating query embeddings. That is,
lines~\ref{line:bwdembed}-\ref{line:bwdcontext2} are removed from
Algorithm~\ref{alg:context}.

Our intuition behind this model is to restrict the search space during
model inference, aiming to reduce the complexity of optimization
compared to the full context model\ignore{ has a larger search space
  by optimizing a sequence of program prediction tasks at one
  time. The amount and quality of training data would be a constraint
  to effectively learn the more complex model in a larger search
  space}. Whether this reduction in optimization complexity is
beneficial to the prediction task is an empirical question we study in
the following section.

\section{Experimental Setup}

{\bf Data Preparation.}  We collected raw data from voice queries
submitted to Xfinity X1 voice-enabled remote controllers during
the week of Feb.\ 22 to 28, 2016.  The dataset contains 32.3M queries
from 2.5M unique viewers. Based on preliminary analyses, we selected
45 seconds as the threshold for dividing successive queries into
sessions, yielding 20.0M sessions in total.

In order to build a training set for supervised learning, we need to
acquire the true user intent for each session. We automatically
extracted noisy labels by examining what the viewer watched after the
voice session. This idea exactly parallels inferring user intent from
clickthrough data in the web domain. No doubt that for web search, the
process of gathering labels is highly refined given the amount of
effort invested by commercial web search engine companies.  By
comparison, our heuristics may seem relatively crude, but given the
paucity of work in this domain, they represent a good initial attempt
to tackle the problem.

If the viewer began watching a program $p$ at most $K$ seconds after
the last query in the voice session $v$ and kept watching it for at
least $L$ seconds, we label the session with $p$. The selection of $K$
and $L$ represents a balance between the quantity and quality of
collected labels. Reducing $K$ and/or increasing $L$ increases the
confidence in the correctness of collected labels but also reduces the
number of labels we obtain, and vice versa. After some initial
exploration, we set $K$ to a relatively small value (30 seconds) and
$L$ to a relatively large value (150 seconds)---which yields
a good balance between data quantity and quality (based on manual
spot-checking). Using these parameters, we extracted 13.0M session-program
pairs.
Note that in reality viewers navigate with a combination of voice queries and
keypad entry, so it is {\it not} the case that our gathered
sessions reflect only successful voice interactions with our X1 platform.

Without any restriction on these sessions, some voice queries might
reflect arbitrary intent (e.g., ``closed caption on'', ``the square
root of eighty one'', ``change to channel 36'', or even complete
gibberish). In order to limit ourselves to voice sessions with a
single clear intent, we used two heuristic strategies to discard
sessions with multiple or unclear intents.  First, we define a way to
reliably predict whether a query is program-related (i.e., the query
is primarily associated with a TV series, movie, video, or sports
program).  We obtain this from the deployed X1 system, which
categorizes each query into one of many action types.  We say a query
is program-related if it is categorized as one of the
following:\ \{SERIES, MOVIE, MUSICVIDEO, SPORTS\}.  Based on this
knowledge, we restricted our data to sessions in which over 2/3 of
queries are program-related and the final query in the session is also
program-related. Since channel changes are a large portion of the
data, this reduces the number of labeled pairs to 2.1M.

Second, we computed the normalized Levenshtein
distance~\cite{yujian2007normalized} between each query pair in the
session, and kept only those sessions where {\it any} pair of queries
has a distance less than 0.5.  Our goal here is to ensure that there
is at least {\it some} cohesion in the sessions.  This heuristic has a
relatively minor effect:\ the resulting filtered dataset contains
1.96M sessions in total.

From this data, we created five splits:\ a training set used in all
experiments, and two groups of development and test sets. The first
development and test sets contain only single-query sessions, called
SingleDev and SingleTest. These are used to study whether the
context-based models hurt accuracy in sessions without context. The
second group contains only multiple-query sessions (i.e., at least two
queries in each session), called MultiDev and MultiTest. In order to
build the global set of programs $\Phi$, we only kept programs if
there are at least 50 associated sessions in the training set,
yielding 471 programs. In other words, our task is 471-way
classification. Statistics for each of the splits are summarized
in Table~\ref{tab:stat}.

\begin{table}[t]
\centering
\begin{tabular}{lcccc}
\hline
Dataset & \# sessions & \# queries & avg.\ session len. & avg.\ query len. \\
\hline
Train & 126016 & 181058 & 1.44 & 2.34  \\
SingleDev & 24792 & 24792 & 1.00 & 2.40 \\
SingleTest & 24572 & 24572 & 1.00 & 2.36 \\
MultiDev & 28427 & 82828 & 2.91 & 2.30 \\
MultiTest & 28173 & 82272 & 2.92 & 2.30 \\
\hline
\end{tabular}
\vspace{0.1cm}
\caption{Dataset Statistics.}\label{tab:stat}
\end{table}

\begin{table*}[t]
\centering
\hspace*{-6em}
\begin{subtable}{0.45\linewidth}
\begin{tabular}{|l|l|l | lll|}
\hline
\textbf{ID} & \textbf{Model} & \textbf{Query} & \textbf{P@1} & \textbf{P@5} & \textbf{MRR} \\
\hline
\hline
1 & EditDist & - & $0.8148^{7}$ & $0.8520^{7}$ & 0.8354 \\
2 & Deployed X1 & - & $0.8743^{1,7}$ & - & - \\
3 & $\textnormal{SVM}^{\textnormal{rank}}$ & - & $0.9131^{1,2,7}$ & $0.9309^{1,7}$ & $0.9267^{1,7}$ \\
\hline
4 & Basic & char & $0.9438^{\textnormal{1-3,7}}$ & $0.9526^{\textnormal{1,7}}$ & $0.9617^{\textnormal{1,3,7}}$ \\
5 & Basic & word & $0.9434^{\textnormal{1-3,7}}$ & $0.9526^{\textnormal{1,7}}$ & $0.9615^{\textnormal{1,3,7}}$ \\
6 & Basic & comb & $\boldsymbol{0.9466^{\textnormal{1-3,7}}}$ & $\boldsymbol{0.9551^{\textnormal{1,7}}}$ & $0.9637^{\textnormal{1,3,7}}$ \\
\hline
7 & Context-f & char & 0.7526 & 0.7936 & 0.8371 \\
8 & Context-f & word & $0.9262^{\textnormal{1,2,7}}$ & $0.9416^{\textnormal{1,7}}$ & $0.9590^{\textnormal{1,3,7}}$ \\
9 & Context-f & comb & $0.9315^{\textnormal{1,2,7}}$ & $0.9474^{\textnormal{1,7}}$ & $\boldsymbol{0.9669^{\textnormal{1,3,7}}}$ \\
\hline
10 & Context-c & char & $0.9378^{\textnormal{1-3,7}}$ & $0.9499^{\textnormal{1,7}}$ & $0.9626^{\textnormal{1,3,7}}$ \\
11 & Context-c & word & $0.9428^{\textnormal{1-3,7}}$ & $0.9502^{\textnormal{1,7}}$ & $0.9608^{\textnormal{1,3,7}}$ \\
12 & Context-c & comb & $0.9435^{\textnormal{1-3,7}}$ & $0.9532^{\textnormal{1,7}}$  & $0.9627^{\textnormal{1,3,7}}$ \\
\hline
\end{tabular}
\caption{single-query}
\label{tab:resultsa}
\end{subtable} 
\begin{subtable}{0.45\linewidth}
\begin{tabular}{|l|l|l | llll|}
\hline
\textbf{ID} & \textbf{Model} & \textbf{Query} & \textbf{P@1} & \textbf{P@5} & \textbf{MRR} & \textbf{QR} \\
\hline
\hline
1 & EditDist & - & 0.4708 & 0.5297 & 0.5033 & 0.834 \\
2 & Deployed X1  & - & 0.4544 & - & - & - \\
3 & $\textnormal{SVM}^{\textnormal{rank}}$ & - & $0.5280^{1,2,7}$ & $0.5985^{1,7}$ & $0.5606^{1,7}$ & $0.949^{1,7}$ \\
\hline
4 & Basic & char & $0.6052^{\textnormal{1-3,7}}$ & $0.6471^{\textnormal{1,3,7}}$ & $0.6901^{\textnormal{1,3,7}}$ & $1.108^{\textnormal{1,3,7}}$ \\
5 & Basic & word & $0.6085^{\textnormal{1-3,7}}$ & $0.6437^{\textnormal{1,3,7}}$ & $0.6773^{\textnormal{1,3,7}}$ & $1.086^{\textnormal{1,3,7}}$ \\
6 & Basic & comb & $0.6135^{\textnormal{1-3,7}}$ & $0.6510^{\textnormal{1,3,7}}$ & $0.6868^{\textnormal{1,3,7}}$ & $1.113^{\textnormal{1,3,7-9}}$ \\
\hline
7 & Context-f & char & 0.4818 & 0.5316 & $0.5803^{1}$ & 0.856 \\
8 & Context-f & word & $0.5989^{\textnormal{1-3,7}}$ & $0.6384^{\textnormal{1,3,7}}$ & $0.6868^{\textnormal{1,3,7}}$ & $1.075^{\textnormal{1,3,7}}$ \\
9 & Context-f & comb & $0.5982^{\textnormal{1-3,7}}$ & $0.6428^{\textnormal{1,3,7}}$ & $0.6883^{\textnormal{1,3,7}}$ & $1.039^{\textnormal{1,3,7}}$ \\
\hline
10 & Context-c & char & $0.6394^{\textnormal{1-9}}$ & $0.6842^{\textnormal{1,3-9}}$ & $0.7306^{\textnormal{1,3-9}}$ & $1.117^{\textnormal{1,3,5,7-9}}$ \\
11 & Context-c & word & $0.6387^{\textnormal{1-9}}$ & $0.6826^{\textnormal{1,3-9}}$ & $0.7290^{\textnormal{1,3-9}}$ & $1.112^{\textnormal{1,3,7-9}}$ \\
12 & Context-c & comb & $\boldsymbol{0.6427^{\textnormal{1-9}}}$ & $\boldsymbol{0.6872^{\textnormal{1,3-9}}}$  & $\boldsymbol{0.7343^{\textnormal{1,3-9}}}$ & $\boldsymbol{1.128^{\textnormal{1,3-9}}}$ \\
\hline
\end{tabular}
\caption{multiple-query}
\label{tab:resultsb}
\end{subtable} 
\caption{Model effectiveness on single-query (left) and multiple-query
  (right) sessions. The second column denotes the model:\ baselines
  compared to the basic, full context (Context-f) and constrained
  context (Context-c) models. The third column indicates the query
  representation. Remaining columns show evaluation metrics. 
  Superscripts indicate the row indexes from which the metric
  difference is statistically significant at $p < 0.01$. Rows are
  numbered in the first column for convenience.}
\label{tab:results}
\end{table*}

\smallskip \noindent \textbf{Model Training}. In total, we have three
options for query representation, \emph{char}, \emph{word},
\emph{combined} (Section~\ref{sec:query}), and three options for the
model, \emph{basic}, \emph{full context}, or \emph{constrained
  context} (Sections~\ref{sec:basic}-\ref{sec:c-context}). Therefore,
we have a total of nine experimental settings, by crossing the three
representations with the three models.

The entire dataset contains 80 distinct characters in total, which
means the size of the one-hot vector used in the \emph{char} setting
is 80. For the word representation, we used 300-dimensional
GloVe~\cite{pennington2014glove} word embeddings to encode each word,
which is trained on 840 billion tokens and freely available. The word
vocabulary of our dataset is 20.4K, with 1759 words not found in the
GloVe word embeddings. Unknown words were randomly initialized with
values uniformly sampled from [-0.05, 0.05].

During training, we used the stochastic gradient descent algorithm
together with $\textrm{RMS-PROP}$~\cite{tieleman2012lecture} to
iteratively update the model parameters. The learning rate was
initially set to $10^{-3}$, and then decreased by a factor of three
when the development set loss stopped decreasing for three epochs. The
maximum number of training epochs was 50. For the constrained context
model, the number of pre-train epochs was selected as 15. The output
size of the LSTMs was set to 200 and the size of linear layer was set
to 150. The regularization weight $\lambda$ was chosen as
$10^{-4}$. At test time, we selected the model that obtained the
highest P@1 accuracy on the development set for evaluation. Our models
were implemented using the Torch framework. We ran all experiments on
a server with two 8-core processors (Intel Xeon E5-2640 v3 2.6GHz) and
1TB RAM, with each experiment running on 6 CPU threads.

\smallskip \noindent \textbf{Baselines.} We considered three baselines
for comparison. The first baseline is the edit distance algorithm, in
which we compared each candidate program's title to the issued query
and returned the program with the smallest edit distance to the query
as the predicted label. Second, we obtained responses from our
production X1 system, which combines statistical machine learning
models with hand-crafted rules to produce the best response.

Third, we built a learning-to-rank baseline using the
$\textnormal{SVM}^{\textnormal{rank}}$ library.  We first used the
edit distance baseline and tf-idf algorithm to find the top 10 closest
programs, and merged them as ranking candidates. We then designed two
types of features:\ (1) the edit distance and tf-idf score between the
query and the candidate programs, and (2) we first computed cosine
similarities of all pairs of word vectors between query and candidate
programs (word vectors were initialized from GloVe embeddings), and then we took the
maximum/mean/minimum values of the word pair similarities as features.

\smallskip \noindent \textbf{Evaluation Metrics.} We used four metrics
to evaluate our models:\ precision at one (P@1), precision at five
(P@5), Mean Reciprocal Rank (MRR), and Query Reduction (QR). The first
three are standard retrieval metrics that are averaged over all
queries, but the last requires some explanation. Query reduction is a
measure of how many queries a viewer has ``saved''. For a session with
$n$ queries, the number of reductions is $n-i$ if the model
returns the correct prediction at the $i$-th query, which means that
the viewer does not need to issue the next $n-i$ queries, hence a
\emph{reduction} of $n-i$. We average this metric over all
sessions. Note that QR is not applicable to single-query sessions.

\begin{figure*}[t]
\centering
\begin{subfigure}{0.24\textwidth}
	\centering
	\includegraphics[width=1.0\linewidth]{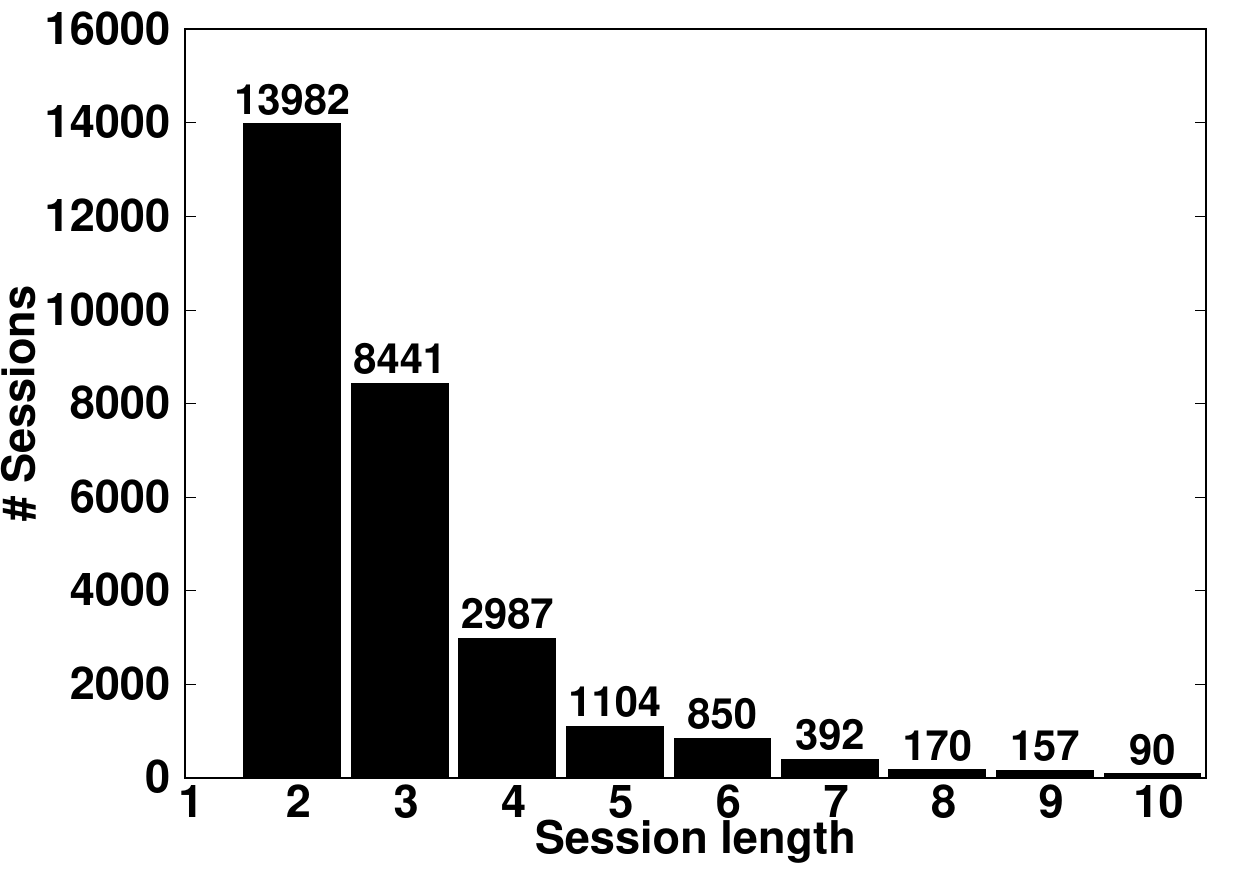}
	\caption{Session Length Distribution}
\end{subfigure}
\begin{subfigure}{0.24\textwidth}
	\centering
	\includegraphics[width=1.0\linewidth]{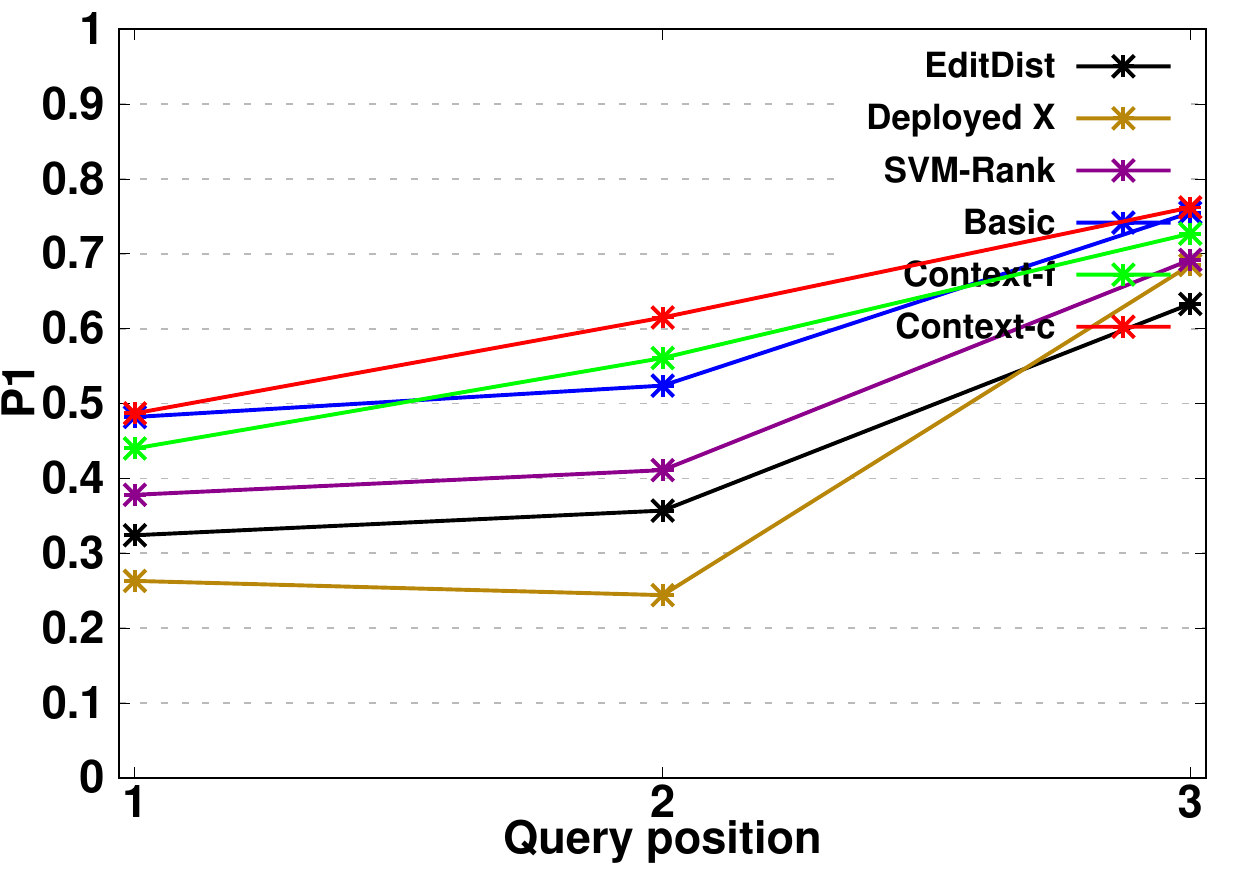}
	\caption{Short Sessions}
\end{subfigure}
\begin{subfigure}{0.24\textwidth}
	\centering
	\includegraphics[width=1.0\linewidth]{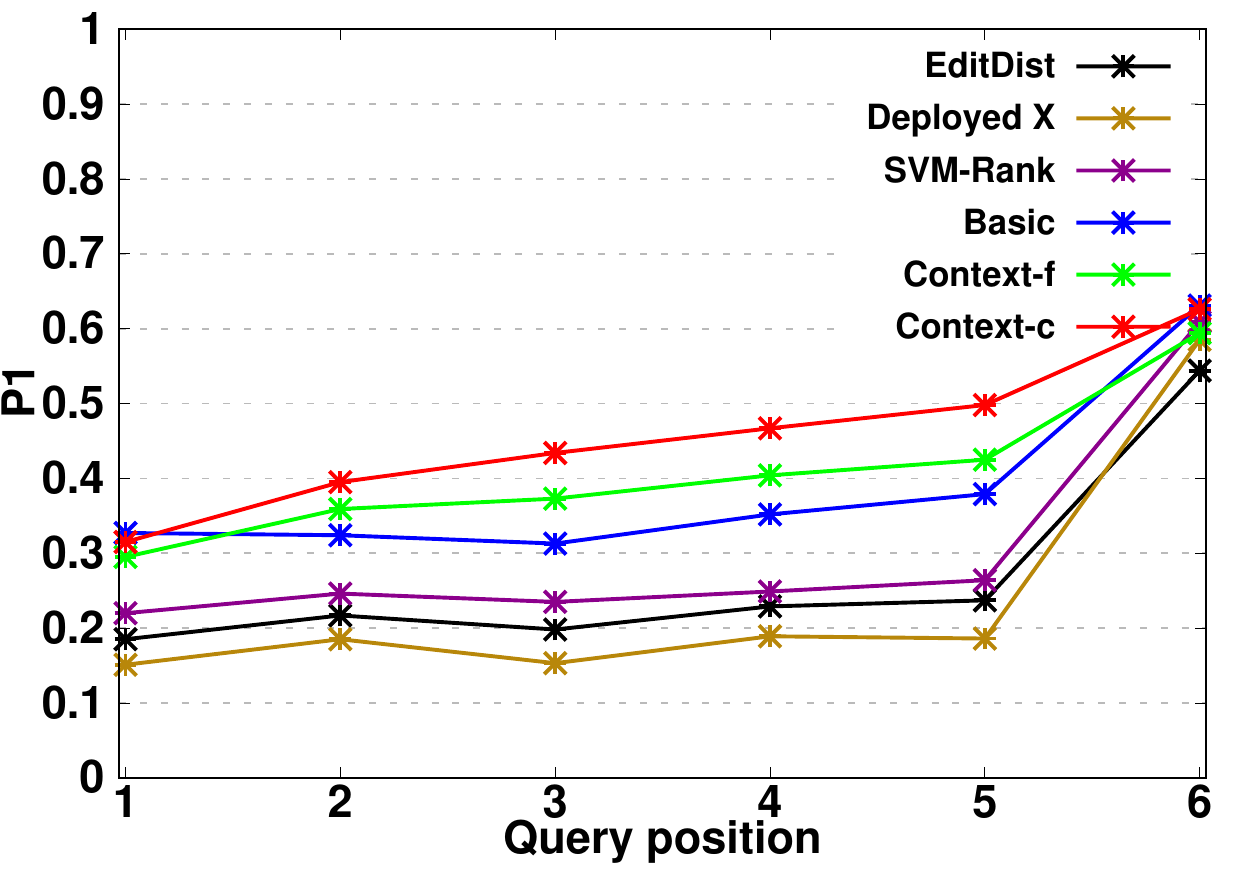}
	\caption{Medium Sessions}
\end{subfigure}
\begin{subfigure}{0.24\textwidth}
	\centering
	\includegraphics[width=1.0\linewidth]{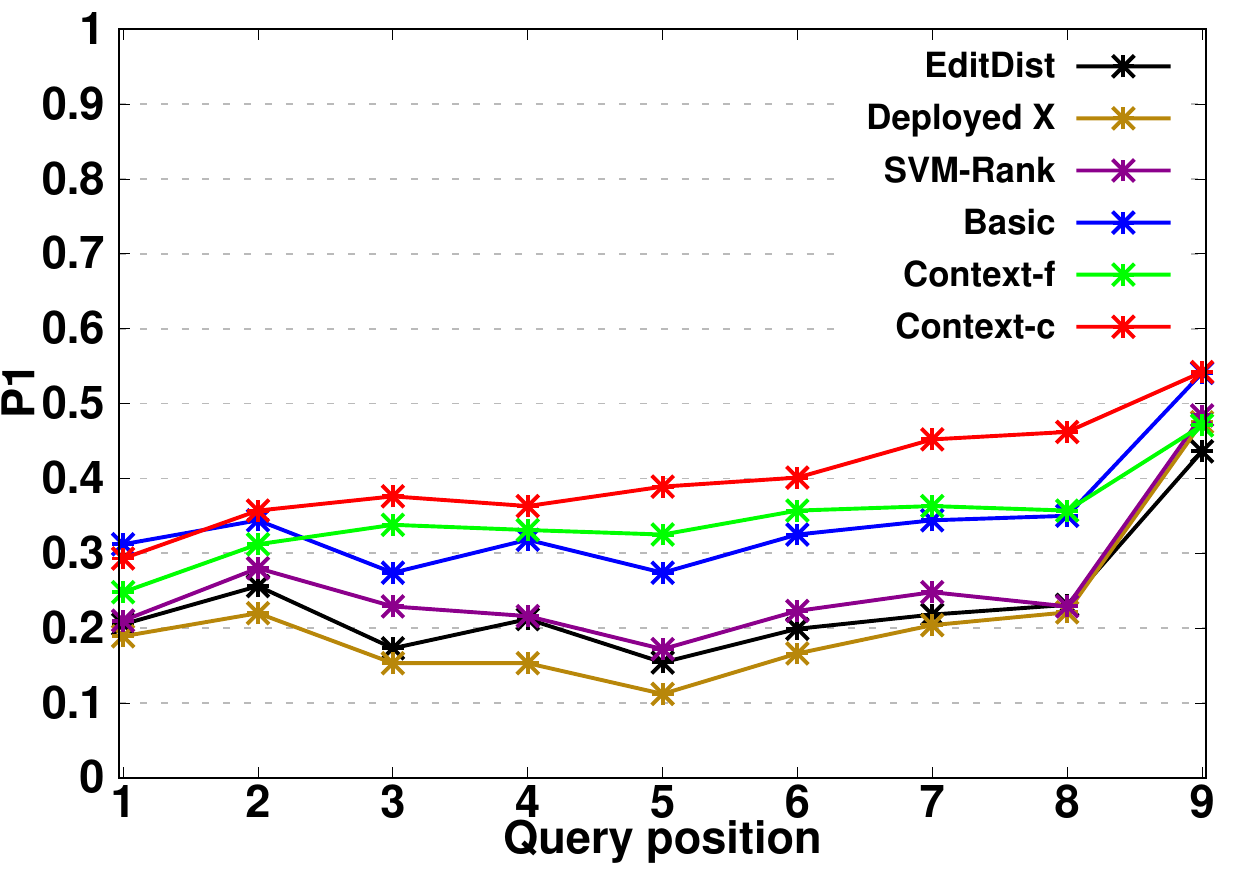}
	\caption{Long Sessions}
\end{subfigure}
\caption{Context analysis of the multiple-query session (MultiTest) test set. The leftmost plot shows the distribution of session lengths. Subfigures (b)-(d) show the average P@1 score at different positions (i.e., the $i$-th query) in the session.}
\label{fig:context-analysis}
\end{figure*}

\section{Results}

Results for the single-query and multiple-query sessions, on the
SingleTest and MultiTest splits, respectively, are shown in
Table~\ref{tab:results}. Each row represents an experimental condition
(numbered for convenience); the second column specifies the model
condition:\ ``Context-f'' denotes the \emph{full} context model and
``Context-c'' represents the \emph{constrained} context model. The
third column indicates the query representation, and the remaining
columns list the various evaluation metrics. Superscripts indicate the
row indexes for which the metric difference is statistically
significant ($p < 0.01$) based on Fisher's two-sided, paired
randomization test~\cite{smucker2007comparison}.  A dash symbol ``-''
connecting two integer indices ``$a$-$b$'' is shorthand for $a, a+1,
\ldots, b-1, b$.

Let's first consider the baselines:\ the current deployed X1 system
achieves fairly high accuracy (P@1 of 0.8743) on single-query
sessions. Since our internal APIs only return the top prediction, we
cannot compute P@5 or MRR. The good accuracy of X1 suggests that
viewers are already fairly satisfied with the current deployed system,
since for single-query sessions they reach their intended programs in
a single shot. In a sense, this is not surprising because, by
definition, these are the ``easy queries''.  The edit distance
baseline also performs fairly well for these easy cases.  The
$\textnormal{SVM}^{\textnormal{rank}}$ predictor achieves better
accuracy than the deployed system because it takes advantage of word
embeddings to consider semantic relatedness at both the character and
the word levels in a supervised setting. Thus, our learning-to-rank
approach forms a reasonably strong baseline.

Turning our attention to baselines on the multiple-query sessions in
Table~\ref{tab:results}(b), we see that accuracy drops significantly
for all baselines. These sessions represent information needs that
weren't satisfied in a single shot, which of course makes them more
challenging. We note that the accuracy of the deployed X1 system
falls below even that of edit distance, which is not surprising as
there would not have been multiple queries if X1 provided the correct
response on the first try.

There are two important questions our experiments are designed to
answer:\ First, which is the most effective query representation
(character, word, or combined)? And second, which is the most
effective context model (no context, full, or constrained)?

For the first question, we observe that in the basic and constrained
context model, the word-level query representation is quite close to
the character-level query representation. However, in nearly all
conditions, across nearly all metrics, the combined condition further
improves (albeit only slightly) upon both representations, which shows
that character-level and word-level representations provide signals
that supplement each other.

In terms of the context models, it seems clear that the constrained
context model significantly outperforms all other models, including
the basic and full context models. Considering that the constrained
context model copies its query embedding layer from the basic model,
we conclude that contextual information does help with the prediction
task. On the other hand, note that the full and constrained models
share exactly the same architecture. The only difference lies in
whether or not we back-propagate to the query embedding layer during
training---the constrained model was designed to restrict the model
search space during model inference. The effectiveness gap on the
multiple-query session dataset demonstrates that the query embedding
layer obtained by the constrained context model through pre-training
is of higher quality. This is likely due to insufficient data for the
full context model to effectively learn parameters for both LSTM
levels. However, a caveat:\ it is conceivable that with even
more training data, the full context model will improve. But
as it currently stands, the constrained context model
displays a better ability to exploit contextual information for
predicting viewers' intent. Overall, for the multiple-query sessions,
the constrained model with the combined representation yields a 41\%
relative improvement over the deployed X1 system for P@1 and a 22\%
relative improvement over $\textnormal{SVM}^{\textnormal{rank}}$.

Looking at the accuracy of the context models for single-query
sessions, we want to make sure that the more sophisticated models do
not ``screw up'' the easy queries. We confirm that this is indeed the
case. It is no surprise that the basic model performs the best on
single-query sessions:\ since there is no context to begin with, all
the ``contextual machinery'' of the richer models can only serve as
distractors. We find that the constrained model with the combined
representation (best condition above) still performs well---slightly
worse (but not significantly so) than the basic model with the combined
representation. It is also interesting to note that the full context
model with the character representation is terrible, which provides
additional evidence that the search space is probably too large with
the combination of much longer query representations and optimization
at the session level.

\begin{table*}
\centering
\hspace*{-2em}
\begin{tabular}{|l|l|l|l|}
\hline
\multicolumn{2}{|l|}{\textbf{Session}} & Cacio : You : You : Caillou & Sienna cover : Color : Casey undercover \\
\multicolumn{2}{|l|}{\textbf{Intended Program}} & Caillou & K.C. Undercover \\
\hline
\hline
\textbf{Model} & \textbf{Query} & \textbf{Example 1} & \textbf{Example 2} \\
\hline
\hline
EditDist & - & $\star$ : House : House : $\star$ & Bee Movie : Room : $\star$ \\
Deployed X1 & - & NA : House : House : $\star$ & NA : In Living Color : $\star$\\
$\textnormal{SVM}^{\textnormal{rank}}$ & - & $\star$ : Now You See Me : Now You See Me : $\star$ & CSI Cyber : Room : $\star$\\
\hline
Basic & char & $\star$ (0.81)  : $\star$ (0.80)  : $\star$ (0.80)  : $\star$ (1.0) &  $\star$ (0.76) : Carolina (0.07)  : $\star$ (0.99) \\
Basic & word & Child Genius (0.03)  : $\star$ (0.57)  : $\star$ (0.57)  : $\star$ (1.0) & Recovery Road (0.48)  : $\star$ (0.08) : $\star$ (0.75) \\
Basic & comb & Paw Patrol (0.17)  : $\star$ (0.83)  : $\star$ (0.83)  : $\star$ (1.0) & $\star$ (0.37) : Magic Mike XXL (0.31) : $\star$ (0.98) \\
\hline
Context-f & char & Lego Ninjago (0.30) : $\star$ (0.79)  : $\star$ (0.90)  : $\star$ (0.99) & $\star$ (0.43) : $\star$ (0.67) : $\star$ (0.89)\\
Context-f & word & Paw Patrol (0.30)  : $\star$ (0.62)  : $\star$ (0.98)  : $\star$ (1.0) & $\star$ (0.29) : $\star$ (0.65) : $\star$ (1.0)\\
Context-f & comb & Lego Ninjago (0.03)  : $\star$ (0.60)  : $\star$ (0.98)  : $\star$ (1.0) & $\star$ (0.41) : $\star$ (0.54) : $\star$ (0.99)\\
\hline
Context-c & char & $\star$ (0.96)  : $\star$ (0.99)  : $\star$ (0.99)  : $\star$ (1.0) & $\star$ (0.81) : $\star$ (0.96)  : $\star$ (0.99)\\
Context-c & word & Wallykazam (0.07)  : $\star$ (0.59)  : $\star$ (0.86)  : $\star$ (1.0) & $\star$ (0.89) : $\star$ (0.80)  : $\star$ (1.0) \\
Context-c & comb & Paw Patrol (0.17)  : $\star$ (0.93)  : $\star$ (1.0)  : $\star$ (1.0) & $\star$ (0.65) : $\star$ (0.83)  : $\star$ (0.97)  \\
\hline
\end{tabular}
\vspace{0.1cm}
\caption{Two sample sessions and top predictions for each model. Each
  query and prediction in the session is separated by a colon. For
  each prediction from our models, we show the confidence
  score. $\star$ indicates that the model response was correct.}
\label{tab:case-study}
\end{table*}

\subsection{Context Analysis}

To better understand how our models take advantage of context, we
focused on multiple-query sessions and examined how accuracy evolves
during the course of a session. Results are shown in
Figure~\ref{fig:context-analysis}. The leftmost plot shows the
histogram of session lengths (i.e., number of queries in a session) in
the MultiTest split (each bar annotated with the actual count). In
Figures~\ref{fig:context-analysis}(b)-(d), we show the average P@1
score from MultiTest at different positions in the session (on the $x$
axis), i.e., at the first query in the session, the second query,
etc. For illustrative purposes we focus on ``short'' sessions with a
length of three (8441 sessions), ``medium'' sessions with a length of
six (850 sessions), and ``long'' sessions with a length of nine (157
sessions). In each plot, we compared our context models with the
baselines. For clarity, in all cases the models used the combined
query representation.

We observe several interesting patterns in
Figures~\ref{fig:context-analysis}(b)-(d). First, for the non-context
models (EditDist, X1, $\textnormal{SVM}^{\textnormal{rank}}$, and
Basic), the accuracy of all queries before the final query is
essentially the same (with small fluctuations due to noise). Accuracy
for the final query rises significantly because the viewer finally
found what she was looking for (and thus is likely to be an ``easy''
query). However, for the context-aware models (Context-f and
Context-c), we observe a consistent increase in the accuracy curves as
the session progresses. This demonstrates that as the model accumulates
more context, it can better identify the viewer's true intent. The
full context model performs consistently worse than the basic model at
the first query, since there is no context. Similarly, the full
context model performs slightly worse than the basic model for the
final query in each session. This finding is consistent with
the results in Table~\ref{tab:results}, since for single-query sessions, the basic model
beats the full context model slightly.

In Table~\ref{tab:case-study}, we provide two real example sessions to
illustrate how each model responds to the sequence of viewer
queries. The session is shown in the first row, where each query is
separated by a colon. The second row shows the viewer's intent (i.e.,
ground truth label). The remaining rows show the output of each model;
due to space limitations, we only show the top predictions along with
their confidence scores for our models. Each prediction in the sequence
is also separated by a colon. To save space, we use the symbol $\star$
to indicate that the prediction is correct.

In the first example (left), the viewer is consistently looking for
the program ``Caillou'', but the query fails three times in a row due
to ASR errors. For the first query ``Cacio'', the edit distance
algorithm can find the intended program because there are many
characters in common. However, X1 failed and labeled this query as NA (i.e., no
answer). For our models, both the Basic/char and Context-c/char models
can predict the correct program from the query ``Cacio'' with high
confidence. However, models with word-level representations all fail
for this query. This is not a surprise as the word ``Cacio'' is a rare
mis-transcription of the word ``Caillou'' and thus rarely seen in the
training set.\footnote{\small ``Cacio'' was found in the GloVe word
  embeddings, thus its word vector was not randomly initialized.} For
the next two successive queries ``You'', all baselines failed. The
basic models still succeed with two identical queries having the same
confidence scores. However, for both the full and constrained context
models, confidence on the second query ``You'' is higher (due to the
previous context). This is an example of how contextual clues can
help, and confirms our intuitions. Since the second example behaves
quite similarly, we omit a description in prose for space
considerations.

\subsection{Manual Evaluation}

As a final summative evaluation to verify our findings, we tested our
models on 100 manually-labeled queries. For these, we randomly
selected queries from multi-query sessions for which the deployed X1
system produced ``no answer'' (NA), which is by construction the most
challenging queries. Results of this evaluation are shown in
Table~\ref{tab:golden}. For brevity, we only examined models with the
combined query representation; ``Prec.''\ indicates P@1 over all
predictions. In this experiment, we allowed the model to not
give an answer by setting a confidence threshold---this allows the model
to trade off coverage and precision. ``Cov.''~indicates
the percentage of queries that yield a response for that threshold,
and ``Prec.'' indicates the precision. For example, with a threshold
of 0.9, the constrained context model can answer 33\% of the queries
at 91\% precision.

We observe that the relative effectiveness of the models is generally
consistent with previous experiments, although for these queries we
see that the full context model beats the basic model.  The
constrained context model is able to correctly respond to about half the
queries that the deployed system completely failed on, which
represents a substantial, real gain. Furthermore, by adjusting the
confidence threshold, we can achieve very high precision at the cost
of coverage. For our best model (constrained context), we can answer
43\% of the queries at 84\% precision.

\begin{table}
\begin{tabular}{| l | l | ll | ll | ll |}
\hline
 & & \multicolumn{2}{c|}{$\boldsymbol{t \geq 0.9}$} & \multicolumn{2}{c|}{$\boldsymbol{t \geq 0.8}$} & \multicolumn{2}{c|}{$\boldsymbol{t \geq 0.7}$}\\
\textbf{Model} & \textbf{Prec.} & \textbf{Cov.} & \textbf{Prec.} & \textbf{Cov.} & \textbf{Prec.} & \textbf{Cov.} & \textbf{Prec.} \\
\hline
\hline
EditDist & 18\% & - & - & - & - & - & - \\
Deployed X1 & 0 & - & - & - & - & - & - \\
$\textnormal{SVM}^{\textnormal{rank}}$ & 29\% & - & - & - & - & - & - \\
\hline
Basic & 38\% & 21\% & 86\% & 24\% & 88\% & 27\% & 85\%\\
Context-f & 45\% & 19\% & 100\% & 25\% & 93\% & 32\% & 88\%\\
Context-c & 50\% & 33\% & 91\% & 38\% & 87\% & 43\% & 84\%\\
\hline
\end{tabular}
\vspace{0.1cm}
\caption{Results on 100 manually-labeled queries. ``Prec.''~indicates
  P@1. We can tradeoff coverage ``Cov.''~with precision~``Prec.''~by
  giving the model the option of not providing an answer, for a
  particular confidence threshold.}
\label{tab:golden}
\end{table} 

\subsection{Efficiency Analysis}

Having obtained significant improvements against strong baselines in
terms of prediction accuracy, we wonder if our neural network models
can achieve sufficiently low latencies for production deployment. To
this end, we studied the training time and test time (i.e., prediction
latency) of our models, shown in Table~\ref{tab:efficiency}. The first two
columns show the experimental setting as before. Column ``\#Params''
shows the total number of parameters in the model, column ``Training''
denotes the training time for each epoch, and column ``Test'' shows
the prediction latency per query. ``Avg.'' indicates the average value
of training/test times, and ``Conf.'' indicates the 95\% confidence
interval of training/test times (both the averaged over 30 epochs).
Overall, the training time of all
models is less than around 100 minutes per epoch, and the per-query
prediction latency is within 8 milliseconds. Most model configurations
converge in the first 20 epochs. This suggests that our models can be
re-trained with a quick turnaround given new data, and that predictions
can be made with low latency. Both are crucial considerations in
production environments.

Comparing the different query representations, we observe that
\emph{combined} has the most number of parameters and was also the
slowest to train and test; \emph{char} has the least number of
parameters but consumed more time in both training and test compared
to the \emph{word} model. For characters, the size of the one-hot
vectors is smaller than that of the word embedding vectors, resulting
in fewer parameters in the character-level LSTM at the query embedding
layer. However, character-level representations are much longer than
word-level representations, which consumes more time when producing
query embeddings.

\begin{table}
\centering
\begin{tabular}{|l|r|r|cc|cc|}
\hline
 &  &  & \multicolumn{2}{c|}{\textbf{Training (min)}} & \multicolumn{2}{c|}{\textbf{Test (ms)}} \\
\textbf{Model} & \textbf{Query} & \textbf{\#Params} & \textbf{Avg.} & \textbf{Conf.} & \textbf{Avg.} & \textbf{Conf.} \\
\hline
\hline
Basic & char & 326,871 & 62.1 & [59.7, 64.7] & 6.4 & [6.2, 6.6] \\
Basic & word & 502,871 & 32.6 & [32.2, 33.0] & 3.0 & [3.0, 3.1] \\
Basic & comb & 758,471 & 94.5 & [91.4, 97.2] & 6.9 & [6.6, 7.0] \\
\hline
Context-f & char & 648,471 & 72.1 & [68.4, 74.5] & 6.6 & [6.4, 7.0] \\
Context-f & word & 824,471 & 58.8 & [57.2, 60.8] & 4.0 & [4.0, 4.1] \\
Context-f & comb & 1,210,071 & 102.4 & [100.8, 103.8] & 7.0 & [6.8, 7.2] \\
\hline
Context-c & char & 648,471 & 32.1 & [30.9, 33.7] & 6.6 & [6.4, 6.8] \\
Context-c & word & 824,471 & 30.1& [29.5, 31.2] & 4.0 & [3.9, 4.1] \\
Context-c & comb & 1,210,071 & 42.5 & [41.8, 43.1] & 6.9 & [6.8, 7.0] \\
\hline
\end{tabular}
\vspace{0.1cm}
\caption{Model efficiency comparisons. Column ``Training'' denotes the
  training time for each epoch, and column ``Test'' shows the
  prediction latency per query. ``Avg.'' indicates the average value
  of training/test times, and ``Conf.'' indicates the 95\% confidence
  interval of training/test times.}
\label{tab:efficiency}
\end{table} 

With the same query representation, the full context models have more
parameters and took longer to train and test than the corresponding
basic models. The extra parameters and training/test latencies come from
the contextual LSTM layer. The constrained context models have the
same number of parameters and similar prediction latencies as the full
context models since they share the same architecture. However, the
training time of the constrained context model is less than half of
the full context models, suggesting that most of the training effort is
spent on the query embedding layer in the full context models.

\begin{figure}
\centering
\begin{subfigure}{0.23\textwidth}
	\centering
	\includegraphics[width=1.0\linewidth]{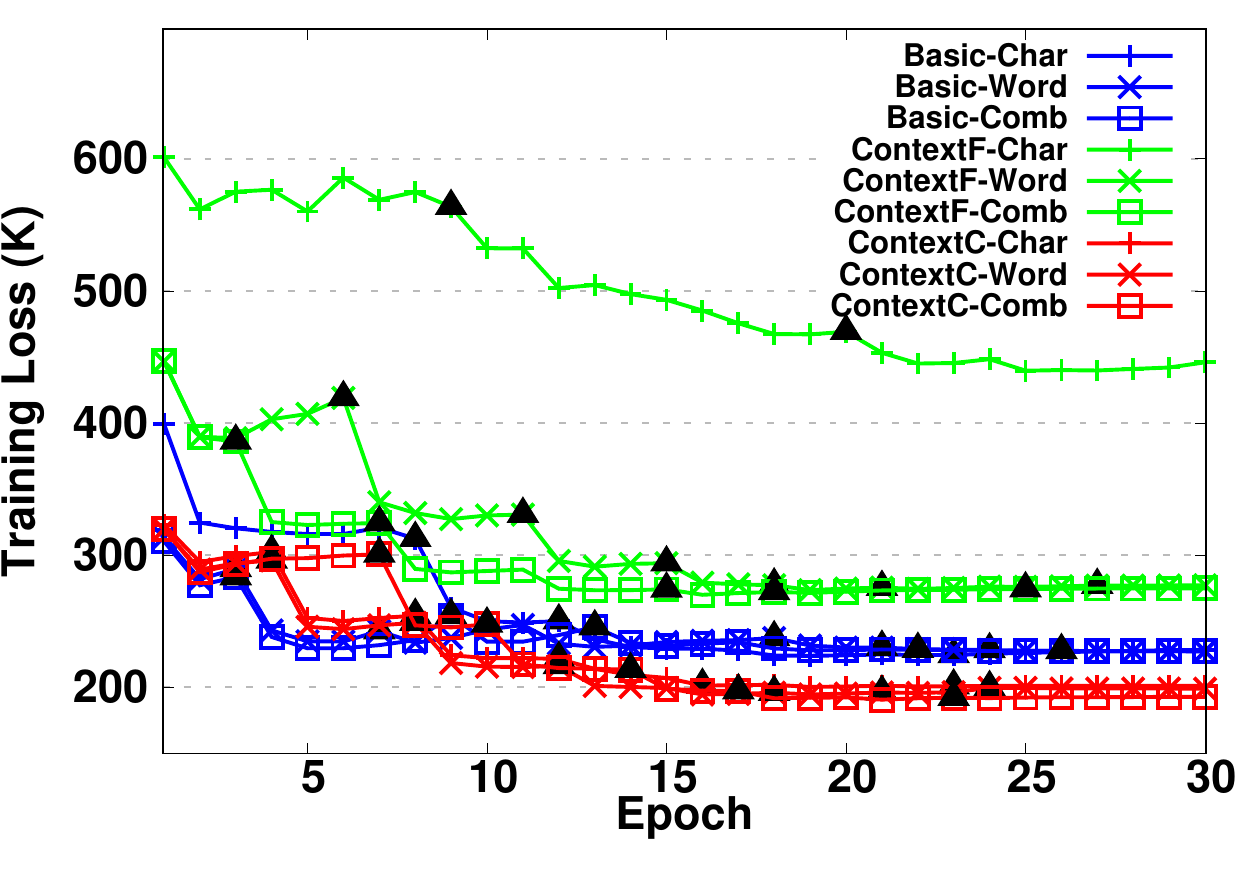}
	\caption{Training Loss}
\end{subfigure}
\begin{subfigure}{0.23\textwidth}
	\centering
	\includegraphics[width=1.0\linewidth]{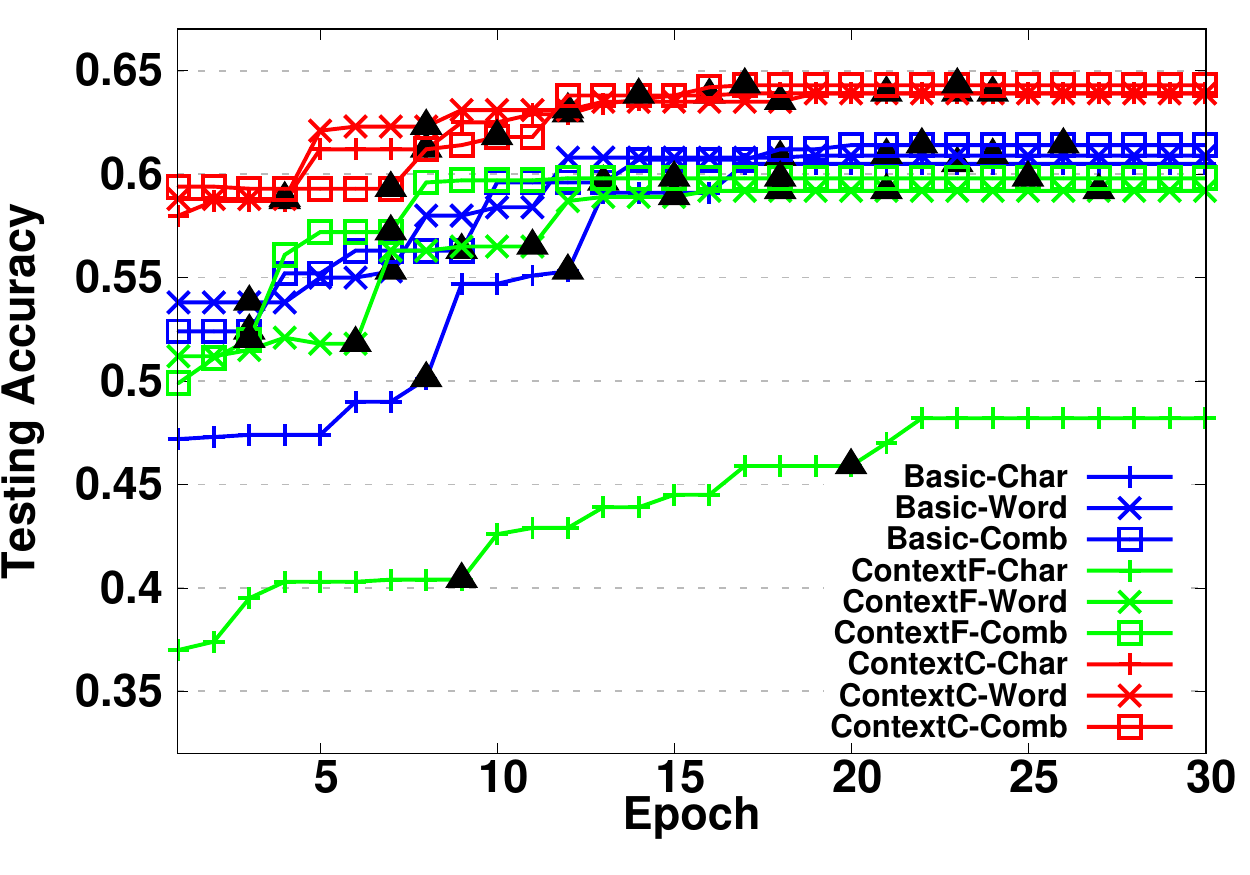}
	\caption{Testing Accuracy}
\end{subfigure}
\caption{Training loss and testing accuracy for each epoch;
  $\blacktriangle$ denotes epochs where the learning rate was
  reduced.}
\label{fig:loss-curves}
\end{figure}

We plot training loss and testing accuracy curves in
Figure~\ref{fig:loss-curves}:\ (a) shows the training loss curve as a function
of epoch, (b) shows the P@1 curve in the MultiTest set
at each epoch. The symbol $\blacktriangle$ denotes
the epochs where the learning rate was reduced by three because
development loss had not decreased for the three epochs. We see
that most models converged within 20 epochs. In the basic and full
context models, the \emph{char} representation took longer to converge
than \emph{word} or \emph{combined}, which shows that a
character-level representation is more difficult to learn. It is also
interesting that the gap in training loss between the basic and full
context models is larger than the gap in test accuracy, which means
that although the full context model is difficult to train, the
benefit of context enables it to generalize well. The constrained
context model walks a middle ground in terms of model complexity and
the ability to capture context information, leading to both lower
training loss and higher test accuracy.

\section{Conclusion}

Our vision is that future entertainment systems should behave like
intelligent agents and respond to voice queries. As a first step, we
tackle a specific problem, voice navigational queries, to help
users find the program they are looking for. We articulate the
challenges associated with this problem, which we tackle with two
ideas:\ by combining word- and character-level representations of
queries, and by modeling session context, both using
hierarchically-arranged neural network modules. Empirically results on a large
real-world voice query log show that our techniques can effectively
cope with ambiguity and compensate for underlying ASR errors. Indeed,
we allow viewers to talk to their TVs, and for customers who learn of
this feature for the first time, it is a delightful experience!



\end{document}